\documentclass[aps,prl,showpacs,notitlepage,singlecolumn,superscriptaddress,11pt,floatfix]{revtex4-2}
\usepackage{amsmath}
\usepackage{amssymb}
\usepackage{amsfonts}
\usepackage{stmaryrd}
\usepackage{graphicx}
\usepackage{textcomp}
\usepackage{url}
\usepackage[colorlinks=true,citecolor=blue,urlcolor=blue]{hyperref}
\usepackage{romannum}
\usepackage{mathtools}
\usepackage[utf8]{inputenc}
\usepackage{braket}
\usepackage{amsmath}
\usepackage{xcolor}
\usepackage{color,soul} 
\usepackage{bm}
\usepackage[normalem]{ulem}
\usepackage[utf8]{inputenc}
\usepackage{array}
\usepackage{makecell}
\usepackage{kotex}
\usepackage{ulem}

\definecolor{mygreen}{RGB}{3,115,45}
\definecolor{mymagenta}{RGB}{186, 00, 132}
\definecolor{mysky}{RGB}{3, 45, 115}

\hypersetup{colorlinks=true, urlcolor=blue, citecolor=cyan, pdfborder={0 0 0}}

\newcommand{\red}[1]{{\color{black}{{#1}}}} 
\newcommand{\ykcc}[1]{{\color{black}{{#1}}}} 

\newcolumntype{L}[1]{>{\raggedright\arraybackslash}p{#1}}
\newcolumntype{C}[1]{>{\centering\arraybackslash}p{#1}}
\newcolumntype{R}[1]{>{\raggedleft\arraybackslash}p{#1}}
\usepackage{multirow}
\usepackage{enumitem}

\usepackage{fancyhdr}
\usepackage{pdfpages} 
\usepackage{pgffor} 

\usepackage[misc,geometry]{ifsym} 

\makeatletter
\AtBeginDocument{\let\LS@rot\@undefined}
\makeatother

\begin{document}
\pagenumbering{arabic}

\title{Coexisting Kagome and Heavy Fermion Flat Bands in YbCr$_6$Ge$_6$}

\author{
Hanoh Lee$^{1,\textcolor{red}{*}}$,
Churlhi Lyi$^{1,\textcolor{red}{*}}$, 
Taehee Lee$^{1}$,
Hyeonhui Na$^{1}$, 
Jinyoung Kim$^{2}$,
Sangjae Lee$^{3}$, 
Younsik Kim$^{2}$,
Anil Rajapitamahuni$^{4}$, 
Asish K. Kundu$^{4}$, 
Elio Vescovo$^{4}$, 
Byeong-Gyu Park$^{5}$,
Changyoung Kim$^{2}$,
Charles H. Ahn$^{3, 6}$,
Frederick J. Walker$^{3}$,
Ji Seop Oh$^{7,8}$, 
Bo Gyu Jang$^{9, \textcolor{red}{\text{\Letter}}}$, 
Youngkuk Kim$^{1, \textcolor{red}{\text{\Letter}}}$, 
Byungmin Sohn$^{1, \textcolor{red}{\text{\Letter}}}$, 
and Tuson Park$^{{1}, \textcolor{red}{\text{\Letter}}}$
\\
$^{1}${\it Department of Physics, Sungkyunkwan University, Suwon 16419, Republic of Korea}\\
$^{2}${\it Department of Physics and Astronomy, Seoul National University University, Seoul 08826, Republic of Korea}\\
$^{3}${\it Department of Applied Physics, Yale University, New Haven, Connecticut 06520, USA}\\
$^{4}${\it National Synchrotron Light Source II, Brookhaven National Laboratory, Upton, New York 11973, USA}\\
$^{5}${\it Pohang Accelerator Laboratory, Pohang University of Science and Technology, Pohang 37673, Republic of Korea}\\
$^{6}${\it Department of Physics, Yale University, New Haven, Connecticut 06520, USA}\\
$^{7}${\it Department of Applied Physics, Sookmyung Women’s University, Seoul 04310, Republic of Korea}\\
$^{8}${\it Institute of Advanced Materials and Systems, Sookmyung Women’s University, Seoul 04310, Republic of Korea}\\
$^{9}${\it Department of Materials Science and Engineering, Kyung Hee University, Yongin
17104, Republic of Korea}\\
$^{\textcolor{red}{*}}${\it These authors contributed equally.}\\
\textsuperscript{\textcolor{red}\Letter}
{\it \href{mailto:bgjang@khu.ac.kr}{\textcolor{blue}{$\mathrm{bgjang@khu.ac.kr}$}}}
\\
\textsuperscript{\textcolor{red}\Letter}
{\it \href{mailto:youngkuk@skku.edu}{\textcolor{blue}{$\mathrm{youngkuk@skku.edu}$}}}
 \\
 \textsuperscript{\textcolor{red}\Letter}
{\it \href{mailto:bsohn@skku.edu}{\textcolor{blue}{$\mathrm{bsohn@skku.edu}$}}}
 \\
 \textsuperscript{\textcolor{red}\Letter}
{\it\href{mailto:tp8701@skku.edu}{\textcolor{blue}{$\mathrm{tp8701@skku.edu}$}}}
 \\
}

\date{\today}
\maketitle

\clearpage
\section*{Abstract}
\textbf{
Flat bands, electronic states with nearly dispersionless energy-momentum structure, provide fertile ground for unconventional quantum phases. Recent observations of flat bands at the Fermi level in kagome metals open the possibility of unifying topology and correlation-driven heavy-fermion physics. Here we show that topology and heavy-fermion correlations coexist in the layered kagome metal YbCr$_6$Ge$_6$. At high temperatures, an intrinsic kagome flat band—arising from frustrated hopping on the kagome lattice—dominates the Fermi level. Upon cooling, localized Yb 4$f$-states hybridize with the topological kagome flat bands, transforming this state into momentum-independent Kondo resonance states across the entire Brillouin zone. 
\ykcc{
Topological analysis of the hybridization gaps reveals filling-tunable weak and strong topological Kondo-insulating regimes, and identifies a topological Dirac--Kondo semimetal.
Taken together, these results identify YbCr$_6$Ge$_6$ as a prototype of a topological heavy-fermion system and a platform where geometric frustration, strong correlations, and topology converge, with broad implications for correlated quantum matter.}
}

%

\clearpage

\section*{Introduction}
Kagome lattice systems lie at the forefront of condensed matter physics, providing a fertile platform for discovering emergent quantum phenomena \cite{yin2022topological, wang2023quantum, wang2024topological}. Their unique corner-sharing triangular lattice constrains electron hopping, leading to characteristic electronic structures such as flat bands (FBs), Dirac points (DPs), and saddle points (SPs) \cite{kiesel2012sublattice, kiesel2013unconventional, ye2018massive, kang2020topological, kang2020dirac, liang2021three, li2021dirac, ghimire2020competing, ortiz2020cs, ma2021rare, chen2021roton, kim2023monolayer, kundu2024low}.
Among these features, FBs stand out due to their high density of states, making them susceptible to interaction-driven instabilities and giving rise to a variety of strongly correlated phases, including unconventional superconductivity \cite{volovik2013flat, cao2018unconventional, cao2018correlated, balents2020superconductivity, nunes2020flat, khasanov2024tuning}, charge density waves \cite{wang2023flat, dalal2024flat}, and various magnetically ordered states \cite{mielke1991ferromagnetism, Tasaki1992, mielke1991ferromagnetism, yin2019negative, yin2020quantum, samanta2024emergence, posey2024two}. More recent studies have revealed that FBs are not merely featureless high-density states but can possess nontrivial quantum geometric and topological properties \cite{peotta2015superfluidity, hu2019geometric, xie2020topology, julku2020superfluid, verma2021optical, aoki2020theoretical, herzog2022superfluid, tian2023evidence}, akin to Landau levels yet emerging without an external magnetic field \cite{rhim2020quantum, wang2021exact, hwang2021geometric, ozawa2021relations, ledwith2023vortexability, fujimoto2024higher, torma2023essay}. This realization has sparked intense interest, as such nontrivial FBs provide a fertile setting for novel strongly correlated phenomena, including fractional quantum Hall effects at zero magnetic field \cite{neupert2011fractional, sheng2011fractional, regnault2011fractional, grushin2012enhancing, scaffidi2012adiabatic, parameswaran2013fractional, bergholtz2013topological, kourtis2014fractional, cai2023signatures, xu2023observation, zeng2023thermodynamic, Park2023}. These advancements not only deepen our understanding of strongly correlated physics but also unlock new opportunities for quantum information science \cite{checkelsky2024flat}.

Heavy fermion systems, first discovered in rare-earth and actinide compounds, have long been recognized as a paradigmatic setting for realizing FBs through strong correlations, where Kondo hybridization between localized $f$-electrons and dispersive conduction bands ($f$-$c$ hybridization) gives rise to anomalously large effective masses~\cite{Stewart84HF, Coleman2007HF, Coleman2015HF, Si2010HF, Wirth2016HF, strydom2023review}. This hybridization generates heavy FBs spanning the entire Brillouin zone (BZ), with their flatness rooted in the localized nature of $f$-electrons~\cite{Coleman2007HF, Shim2007HF, Si2010HF, Choi2012HF, Coleman2015HF}. Despite this well-established mechanism, recent breakthroughs in Moir\'e superlattices \cite{park2008new, bistritzer2011moire, dean2013hofstadter, hunt2013massive, cao2018unconventional, tang2020simulation, regan2020mott, xu2020wigner}, where geometric and correlation effects intertwine to produce FBs, have shifted the focus toward their intrinsic topological properties and their connection to quantum geometry \cite{claassen2015position, song2022magic, yu2023magic, herzog2404topological}. This renewed perspective has invigorated efforts to explore how geometric and topological FBs intersect with Kondo-driven topological phenomena \cite{dzero2010topological, dzero2016topological, lai2018weyl, dzsaber2021giant, dzsaber2022control, iraola2024topology, singh2024topological}, collectively referred to as topological heavy fermion systems. However, the interplay between heavy fermion systems and modern geometric FB concepts remains largely uncharted, hindered by the scarcity of material realizations.

Here we report the coexistence of two flat bands near the Fermi level (E$_{\rm F}$) of the kagome compound YbCr$_6$Ge$_6$ (YCG), where geometrically frustrated flat bands (KFBs) originating from the 3$d_{z^2}$ orbitals of Cr~\cite{yang2022fermi, wang2020experimental, guo2024ubiquitous, xu2023frustrated} coexist with Kondo resonance states (KRSs) from Yb $f$-orbitals \red{within the experimental resolution ($\sim$~20~meV)}. Angle-resolved photoemission spectroscopy (ARPES) measurements, complemented by DFT+DMFT calculations, show that the $f$-electrons of Yb, intercalated between Cr-kagome layers, undergo Kondo hybridization with the kagome bands, forming FBs that span the entire BZ. Transport measurements corroborate the Kondo resonance behavior with the coherent temperature ($T_\text{coh}$) of 80~K (See Supplementary Note 1). \red{Our theory shows that crystalline symmetry forbids hybridization along certain high-symmetry lines, stabilizing Dirac crossings of heavy-fermion character. 
%
%
\ykcc{
The Fu--Kane parity analysis of the Kondo-coherent low-energy band structure distinguishes gapped weak and strong topological Kondo-insulating phases and reveals a Dirac--Kondo semimetal stabilized by symmetry-protected band crossings.
}
Hence, YCG serves as a prototype of a topological heavy-fermion system and a platform to explore symmetry-protected quasiparticles in strongly correlated quantum matter.
}

\section*{Results}
\begin{figure} 
    \centering
    \includegraphics[width=\textwidth]{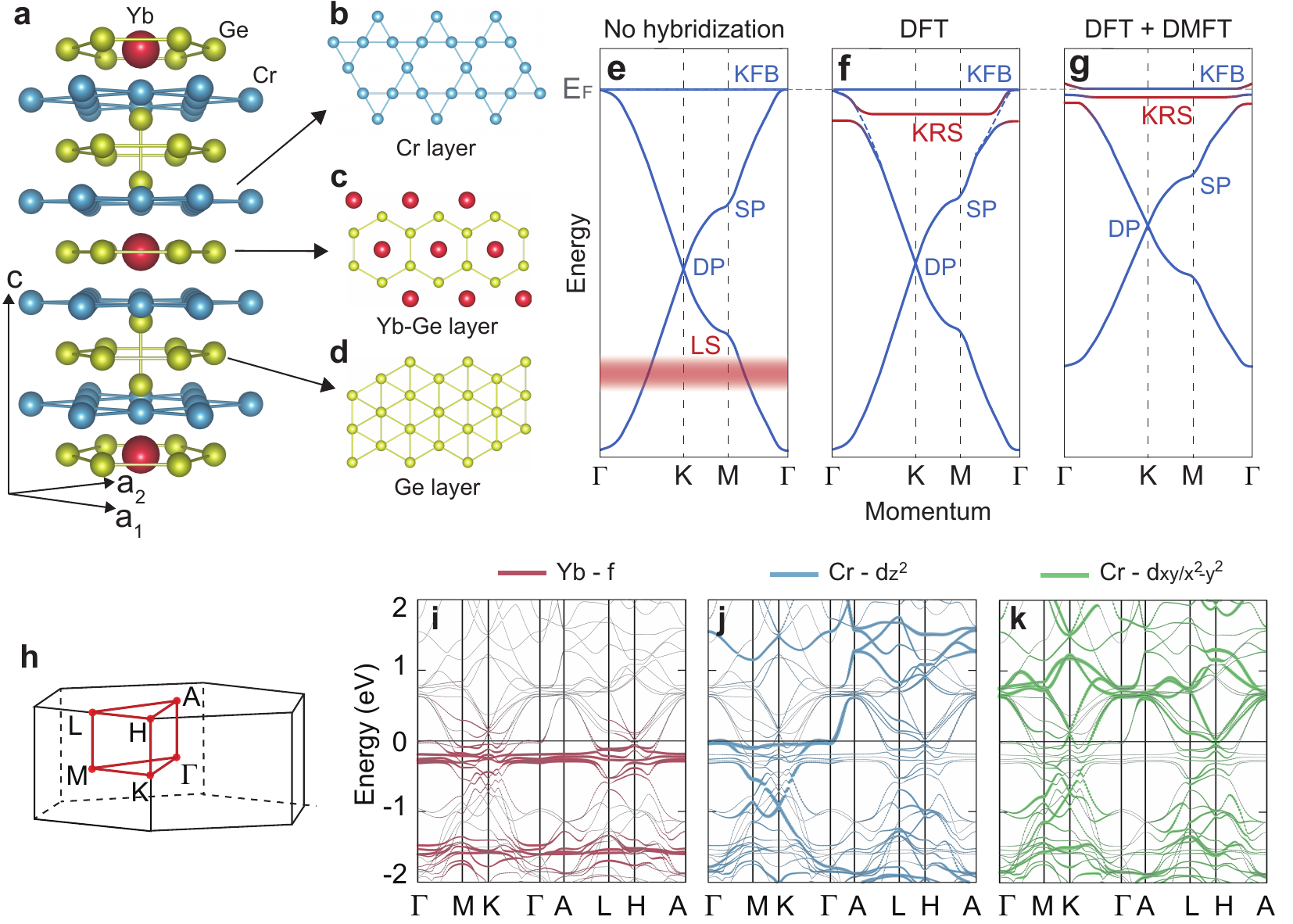}
    \caption{
    {\bf  Atomic and electronic structure of Kondo kagome system YbCr$_6$Ge$_6$ (YCG).}
(a) Atomic structure of the kagome lattice system YCG, delineating the layered arrangement of Yb, Cr, and Ge. 
(b) Kagome layer composed of Cr atoms. 
(c) Honeycomb Ge layer with Yb atoms situated at the hexagonal centers. 
(d) Honeycomb Ge layer with perpendicular Ge dimers at the hexagonal centers. 
(e-g) Schematic illustrations of the Cr kagome bands and localized Yb $f$-orbitals (e) without hybridization, (f) in DFT, and (g) DFT+DMFT, respectively. The Dirac point (DP), van Hove singularities saddle point (SP), and  KFB are highlighted. LS and KRS represent localized states and Kondo resonance states of Yb $f$-orbitals, respectively.
(h) Brillouin zone (BZ) showing the high-symmetry points used in calculations. 
(i-k) DFT Band structures projected onto the $f$ orbitals (red), Cr $3d_{z^2}$ orbitals (blue), and Cr $d_{x^2-y^2}/d_{xy}$ (green), respectively\red{, obtained with $U=0$ for both Cr $d$ and Yb $f$ orbitals}. The zero-energy level is set at the charge-neutral electron filling.
\label{fig1}
}
\end{figure}

The kagome compound YCG crystallizes in a family of kagome layered structure \cite{li2021dirac, ghimire2020competing, ma2021rare} (Fig.\,\ref{fig1}a) with a unit cell composed of four alternating layers of Yb (red), Cr (blue), and Ge (yellow) atoms. At the base is a Cr kagome layer (Fig.\,\ref{fig1}b), followed by a Ge honeycomb layer in which a Yb atom occupies the hexagon center (Fig.\,\ref{fig1}c). This is followed by another Cr kagome layer, capped by a Ge honeycomb layer featuring a perpendicular Ge dimer aligned with the hexagon center (Fig.\,\ref{fig1}d). These four layers repeat periodically along the vertical axis. The alternating sequence of Cr double kagome layers and Ge honeycomb layers preserves the full crystalline symmetry of space group $P6/mmm$ (\#191). Notably, the structure is inversion symmetric—with the inversion center located at either the Yb atom or the Ge dimer—and maintains both planar mirror symmetry ($m_z$) and six-fold rotational symmetry ($C_{6z}$). 

Figures\,1e-1g illustrate the coexistence of KFB and KRS in YCG, providing a schematic overview of the band evolution. The Cr $3d$-orbitals in the kagome layers give rise to characteristic kagome bands, including a FB, a DP, and a SP. In YCG, the KFB is located at $E_{\rm{F}}$, as shown in Fig.\,\ref{fig1}e. When there is no $f$-$c$ hybridization, the Yb $4f$ levels remain far below $E_{\rm{F}}$, corresponding to nearly atomic $f_{7/2}$ states, forming the localized magnetic moments. As $f$-$c$ hybridization takes place, quasiparticle peaks with heavy effective masses develop near $E_{\rm{F}}$. Figure\,1f presents a schematic band structure based on DFT, where the $4f$ states participate in the band formation near $E_{\rm{F}}$ while maintaining their non-dispersive nature. 

When the strong correlation effects of Yb 4$f$ orbitals are properly taken into account, the quasiparticle bands arising from $f$-$c$ hybridization undergo significant renormalization, which leads to the KRS being energetically aligned with the KFB right near $E_{\rm{F}}$ (Fig.\,\ref{fig1}g).
Kondo screening develops as the temperature decreases, leading to renormalized coexisting FB states at $E_{\rm{F}}$, with the FBs extending across the entire BZ.
\red{In addition, overall Cr 3$d$ bands are also renormalized, leading to an overall narrowing of the bandwidth.}

Density functional theory (DFT) calculations reveal that YCG is a KFB system, where the kagome bands derived from Cr $3d_{z^2}$ orbitals dominate the density of states near E$_{\rm F}$, with Yb $4f$ states positioned nearby in energy. Figures\,1i-k display the band structure of YCG using first-principles calculations \red{(See Supplementary Notes~3.4 and 3.6 for detailed analyses of the wave functions and orbital contributions, respectively.)
}. The Cr $d_{z^2}$ orbital contributes to the flat bands near $E_{\rm F}$ (Fig.\,\ref{fig1}j), while the Yb $4f$-states form flat bands located approximately 0.2~–~0.3 eV below E$_{\rm F}$ (See Fig.\,\ref{fig1}i). The localized $f$-orbitals and the KFBs exhibit distinct characteristics: the $f$-orbitals are flat across the entire momentum space, whereas the KFBs are dispersionless only within specific $k_z$ plane. The significant out-of-plane dispersion highlights the role of interlayer interactions, suggesting possible hybridization between the intercalated Yb $4f$-orbitals and the kagome layers (Fig.\,\ref{fig1}e). \red{DFT alone cannot capture the effects of strong electronic correlations properly, and this limitation is remedied in our DFT+DMFT approach as discussed in the following sections.}

Figure\,2a shows the electronic structure of YCG from ARPES measurements, where the $k_x$-$k_y$ constant-energy maps at $E = E_{\rm F}$ and $E = E_{\rm F} - 0.3$ eV reveal two bands crossing $E_{\rm F}$: a hole-like band centered at $\overline{\rm \Gamma}$ (the $\alpha$ band) and the other band centered at $\overline{\rm K}$ (the $\beta$ band). The Fermi momentum $k_{\rm F}$ of the $\beta$ band is located at $\overline{\rm K}$ within our experimental resolution, showing that DP of the $\beta$ band is positioned near $E_{\rm F}$. The $\alpha$ band is barely visible in the first BZ, likely due to matrix element effects~\cite{moser2017experimentalist, kang2020dirac}. \red{Both the $\alpha$ and $\beta$ bands are identified as bulk bands, as they are observed across different surface terminations (See Supplementary Note 2.3).}

\begin{figure}        
    \centering 
    \includegraphics[width=0.9\textwidth]{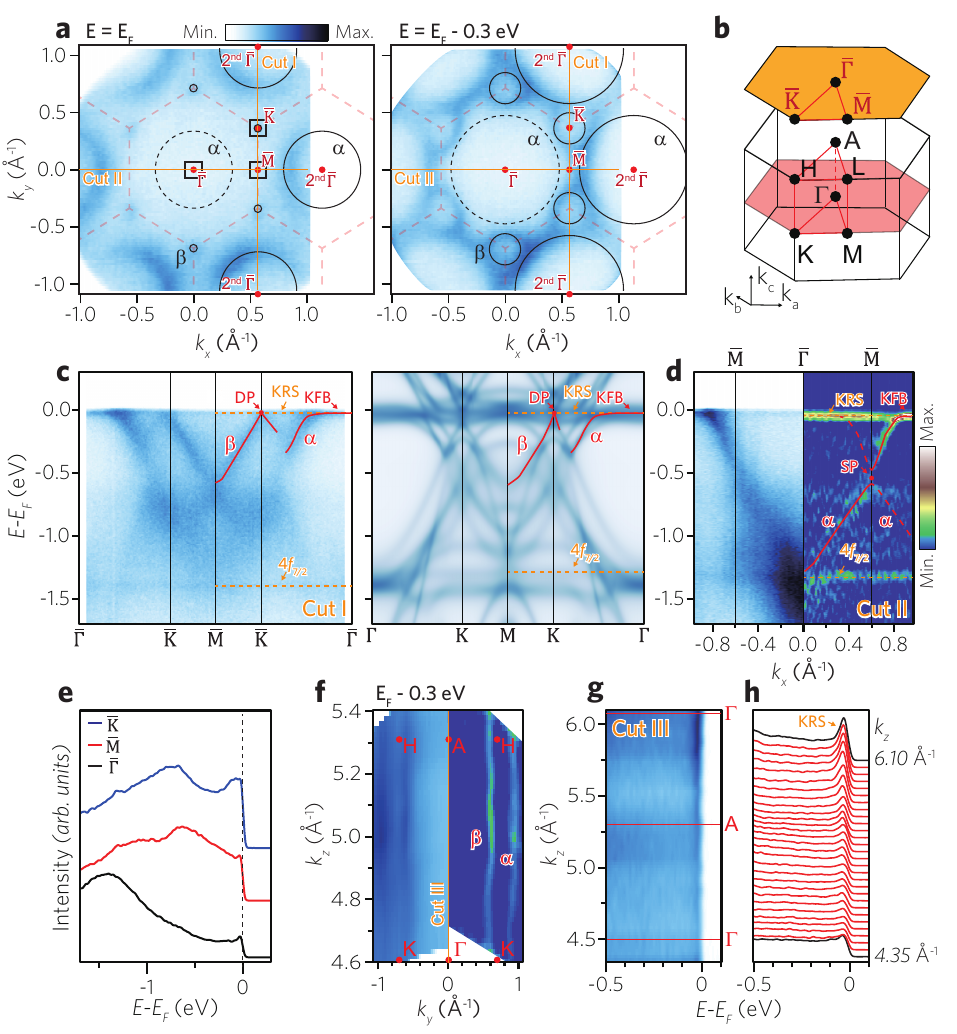}
    \caption{
    {\bf Momentum-resolved electronic structure of YCG measured by ARPES \red{at 18~K}.} 
    (a) $k_{x}$-$k_{y}$ constant-energy maps at the Fermi level (E$_F$) and E~=~E$_F$~-~0.3~eV, showing the $\alpha$ and $\beta$ bands crossing $E_{\rm F}$. 
    (b) Schematic of the 3D BZ with a projection onto the 2D plane.
    (c) Energy-momentum (E-$k$) cut along $\overline{\Gamma}$-$\overline{\rm K}$-$\overline{\rm M} $ (cut I in (a)). DP, KRS, and KFB refer to Dirac point, Kondo resonance state, and kagome flat band, respectively. The $\alpha$ and $\beta$ bands are highlighted with red lines. \red{The corresponding DFT+DMFT-calculated E-$k$ cut is also shown.}
    (d) E-$k$ cut along $\overline{\rm M}$-$\overline{\Gamma}$-$\overline{\rm M}$ (cut II in (a)) with the corresponding 2D curvature intensity plot~\cite{zhang2011precise}. Due to the matrix element effect, parts of the $\alpha$ band are not visible and are represented with red dotted lines.
    (e) Energy distribution curves (EDCs) at $\overline{\Gamma}$, $\overline{\rm K}$, and $\overline{\rm M}$ points \red{extracted from black rectangular regions in panel (a).}
    (f) $k_{y}$-$k_{z}$ constant-energy map at $E$~=~$E_{\rm F}$~-~0.3~eV with the corresponding 2D curvature intensity plot.
    (g) E-$k_{z}$ cut along $\Gamma$-${\rm A}$ (cut III in (e)).
    (h) $k_{z}$-dependent EDCs showing consistently strong peaks at $E_{\rm F}$. 
    }
    \label{fig2}
\end{figure}

The band dispersions observed along high-symmetry directions in YCG clearly reveal the Cr $d$-orbital kagome bands, as predicted by our DFT calculations. Here we annotate high symmetry points using the 2D surface BZ notation as the features are almost $k_z$-independent \red{(Fig.\,\ref{fig2}b)}. Figures \,\ref{fig2}c and \ref{fig2}d present the energy-momentum band dispersions along the $\overline{\Gamma}$-$\overline{\rm K}$-$\overline{\rm M}$ and $\overline{\rm M}$-$\overline{\Gamma}$-$\overline{\rm M}$ high-symmetry lines (Cut I and Cut II), respectively. Notably, at $E = E_{\rm F}$, the $\beta$ band exhibits Dirac-like dispersion at the $\overline{\rm K}$ point. For the $\alpha$ band, the FB and SP appear at $\overline{\rm \Gamma}$ and $\overline{\rm M}$ at $E = E_{\rm F}$ and $E = E_{\rm F} - 0.55$ eV, respectively. The distinct visualization of these features underscores the role of kagome-derived states in shaping the electronic properties of YCG \red{, which are also clearly reproduced in our DFT+DMFT calculations that explicitly account for the correlation effects of both Yb 4$f$ and Cr 3$d$ electrons (Fig.\,\ref{fig2}c) (See Supplementary Note~3.5 for the full DFT+DMFT energy spectrum.). Compared to the DFT results shown in Fig.~\ref{fig1}, the Yb 4$f$ levels are strongly renormalized, giving rise to a pronounced KRS at $E_{\rm F}$. In addition, the Cr $d$ bands are also renormalized and become incoherent.} We note that these kagome-driven band features are also observed in YCr$_6$Ge$_6$~\cite{yang2022fermi,wang2020experimental, Ishii13p023705}.

In addition to the kagome $d$-orbital bands, the KRS can be identified in the ARPES results, originating from Yb $4f$ states as anticipated in DFT calculations. As shown in Fig.\,\ref{fig2}c, this FB is prominently observed near $E_{\rm F}$ and exhibits a defining characteristic: its momentum independence in the $k_{x}$-$k_{y}$ plane (Figs.\,\ref{fig2}c-d), along with its coherence. This is evidenced by strong intensity peaks in the energy distribution curves (EDCs) at high-symmetry points (Fig.\,\ref{fig2}e). Extending this analysis to three dimensions, Fig.\,\ref{fig2}f presents the $k_{y}$-$k_{z}$ constant-energy map at $E = E_{\rm F} - 0.3~$eV \red{(See Supplementary Note 2.4 for additional $k_z$-dependent ARPES results)}. The $E$-$k_{z}$ cut along the $\Gamma$-$\rm A$ high-symmetry line (Cut III in Fig.\,\ref{fig2}f) further illustrates the presence of the KRS at $E_{\rm F}$ throughout the high-symmetry line (Fig.\,\ref{fig2}g). Moreover, the $k_{z}$-dependent EDCs in Fig.\,\ref{fig2}h demonstrate that peaks from the KRS consistently appear at $E_{\rm F}$, regardless of the $k_{z}$ momenta. These observations show that the KRS at $E_{\rm F}$ is a highly localized state, extending across the entire BZ and remaining independent of the electron momenta, $k_{x}$, $k_{y}$, and $k_{z}$. 

\red{
Furthermore, we conduct two additional experiments supporting that the KRS originates from Yb $4f$ states. First, we perform ARPES on LuCr$_6$Ge$_6$ (LCG), where heavy-fermion behavior is absent (see Supplementary Notes 1.1 and 2.1), and confirm that the KRS does not appear in LCG. Second, we perform Yb-resonant photoemission spectroscopy on YCG and observe that the KRS exhibits a pronounced intensity dependence across the Yb N$_5$ edge ($\sim$~183 eV, see Supplementary Note 2.2). These results provide strong evidence that the KRS arises from Yb $4f$ orbitals.
}



\red{One can expect to observe different temperature-dependent behaviors in KRS and KFB, as they originate from different mechanisms.} Figures\,3a–3c show $E$-$k$ cuts along the $\overline{\rm M}$–$\overline{\Gamma}$–$\overline{\rm M}$ high-symmetry line, measured at temperatures of \red{220~K, 80~K, and 18~K, respectively}. Around the first $\overline{\Gamma}$, where the $\alpha$ band is suppressed due to matrix element effect, the KRS is prominently observed at 18~K but progressively diminishes in intensity as the temperature increases. Eventually, at 220~K, the FB around the first $\overline{\Gamma}$ becomes nearly undetectable, while the $\alpha$ band remains visible near the second $\overline{\Gamma}$. This behavior is further illustrated in the energy distribution curves (EDCs) at $\overline{\Gamma}$ (Fig.\,\ref{fig3}d), where the FB peak broadens and vanishes with increasing temperature \red{(See Supplementary Note 2.5 for fitting results)}. The disappearance of the FB peak at higher temperatures suggests possible Kondo behavior, originating from the hybridization between localized Yb 4$f$ orbitals and Cr 3$d$ conduction electrons.  

In stark contrast to the KRS, the KFB from the Cr $3d_{z^2}$ orbitals remains coherent above the Kondo temperature scale. Fig.\,\ref{fig3}e presents the temperature-dependent EDC at $k_x$~=~1.0~$\AA^{-1}$ as a function of temperature. The peaks near $E_{\rm F}$, marked with red inverted triangles, persist up to 220~K, indicating that this feature does not share the same origin as the KRS. Thus, we conclude that the dispersionless feature at $k_x$~=~1.0~$\AA^{-1}$ \red{(near the second $\overline{\Gamma}$)} near $E_{\rm F}$ corresponds to the KFB, which originates from the out-of-plane hybridization of Cr $3d_{z^2}$ orbitals across adjacent kagome layers~\cite{Ishii13p023705}.

Figure\,3g shows the DFT+DMFT spectral function calculated at 58~K, where the FB is observed near $E_{\rm F}$, as a direct consequence of Kondo hybridization. Notably, the interaction between the Cr kagome bands (marked as $d$ in Figs.\,\ref{fig3}f and \ref{fig3}g) and Yb 4$f$ states gives rise to a characteristic hybridization gap. As the temperature increases, these KRSs become incoherent and suppressed. Eventually, the FBs near $E_{\rm F}$ are no longer clearly defined \red{and the hybridization gap features also diminish} at 290~K as shown in Fig.\,\ref{fig3}f. \red{The temperature-dependent features are also well established in Fig.\,\ref{fig3}h.} Unlike the DFT-only results (Fig.\,\ref{fig1}i), where the Yb 4$f$-orbital bands are positioned ∼0.3 eV below E$_{\rm F}$ with an absence of kagome flat bands along the $L$-$A$-$L$ direction at $E_{\rm F}$, the DFT+DMFT calculations properly capture the strong electron correlation effects of the Yb 4$f$ orbitals, yielding strongly renormalized Yb 4$f$ bands that align well with the ARPES observations.

As a consequence of the coexistence of KRS and KFBs near the Fermi energy, we argue that YCG realizes a Dirac--Kondo semimetallic phase, offering a possible manifestation of topological heavy-fermion quasiparticles. 
In this framework, KFBs---localized within the kagome planes but dispersive along the $c$-axis due to interlayer coupling via Yb atoms---participate in the Kondo screening of Yb $4f_{7/2}$ moments, generating momentum-independent Kondo resonance states at low temperature ($T \ll T_K$) (See Figs.\,\ref{fig:kfb-krs}a-b).
Crucially, crystalline symmetry forbids hybridization along certain high-symmetry lines, enforcing symmetry-protected Dirac band crossings with heavy-fermion character, while hybridization gaps open elsewhere. 
These gapless excitations, which retain both strong-correlation coherence and topological protection, are referred to as topological heavy fermions~\cite{dzero2010topological,dzero2016topological,Dzsaber17p246601,dzsaber2021giant,lai2018weyl,Grefe20p2469,feng2016dirac,Chen22p1341}. Moreover, the hybridization-induced gaps that open away from the Dirac points carry nontrivial $\mathbb{Z}_2$ invariants in the corresponding time-reversal-invariant planes of the Brillouin zone, as demonstrated below using DFT calculations and topological analyses.

To demonstrate that YCG realizes a topological heavy-fermion system, we construct an effective DFT band structure that reproduces the DFT+DMFT spectral function and, consistently, the low-temperature ARPES data \red{(See Supplementary Note 3.4 for a direct comparison between the DFT+DMFT and DFT+$U$ band structures.)}. 
Figs.\,\ref{fig:kfb-krs}c-d highlight the contrast between the high- and low-temperature electronic structures. 
Fig.\,\ref{fig:kfb-krs}c shows the band structure of LCG, taken as a reference for the high-temperature limit of YCG \red{(See Supplementary Note\,3.1 for the full energy spectra of YCG and LCG)}. In this case, only kagome-derived states are present and no $f$ states appear near the Fermi level. The kagome bands are clearly resolved on the $k_z=0$ plane, and the nominally flat bands disperse upward with $k_z$, demonstrating noticeable interlayer hopping characteristic of the double-kagome stacking.
In contrast, Fig.\,\ref{fig:kfb-krs}d (YCG), obtained in an effective DFT calculation with $U_{\mathrm{Cr}\,d}=2.6$~eV and $U_{\mathrm{Yb}\,f}=0$~eV, captures the KRS–KFB hybridization and reproduces the DFT+DMFT/ARPES results; the eight branches of the Yb $4f_{7/2}$ manifold (four Kramers-degenerate pairs) overlap the kagome flat bands.

\ykcc{
The emergence of nontrivial $Z_2$ topology in multiple hybridization gaps demonstrates that
topological heavy-fermion quasiparticles are stabilized by the interplay between KRSs and
KFBs over a finite range of chemical potential near the Fermi level rather than requiring
fine tuning to a single filling.
This conclusion is substantiated by a Fu--Kane parity analysis of the hybridization-induced
gaps highlighted in Fig.~\ref{fig:kfb-krs}d
(Figs.~\ref{fig:kfb-krs}e--g).
When the chemical potential is fixed to the experimentally determined Fermi level, the
low-energy Kondo-hybridized quasiparticle spectrum encounters  three distinct low-energy regions, which are marked by red, green, and blue shading in Fig.~\ref{fig:kfb-krs}d.
The red-shaded gap realizes a weak topological Kondo insulator (TKI) with
$(\nu_0;\nu_1\nu_2\nu_3)=(0;001)$ and therefore
$\nu_{2\mathrm{D}}(k_z=0)=\nu_{2\mathrm{D}}(k_z=\pi)=1$,
whereas the green-shaded gap corresponds to a strong TKI characterized by
$(\nu_0;\nu_1\nu_2\nu_3)=(1;001)$.
By contrast, the blue-shaded region realizes a topological Dirac--Kondo semimetal (DKSM) because the
$k_z=\pi$ plane hosts a nontrivial $Z_2$ invariant while symmetry-protected DPs persist
along the high-symmetry lines $\Gamma$--$A$ and $K$--$H$, which prevents the opening of a
fully gapped quasiparticle spectrum.
The protection of these DPs follows directly from symmetry since along $\Gamma$--$A$ the
flat $f$-derived states transform as $\overline{\Delta}_9$ whereas the dispersive kagome
states transform as $\overline{\Delta}_7$, which forbids their hybridization and stabilizes
the Dirac points. Analogous crossings arise along $K$--$H$ from the crossing between
$\overline{P}_4\oplus\overline{P}_5$ and $\overline{P}_6$ bands
(See Supplementary Notes~3.2 and 3.3 for detailed topological analysis).

Several remarks on the filling-tunable topological phases in YCG, including the DKSM regime, help clarify the robustness and implications of our prediction.
}
The effective DFT bands, constructed in the zero-temperature quasiparticle limit, qualitatively reproduce the DFT+DMFT spectral function and ARPES data. Importantly, not only the Yb $f$ states but also the Cr $d$ orbitals require on-site interactions to reproduce the ARPES energy spectrum, indicating that strong correlation effects are essential for the kagome flat bands as well. 
\ykcc{
Although static DFT does not capture the full renormalization of the KFBs and KRSs, the central
conclusions remain robust. Crystalline symmetry forbids hybridization along specific high-symmetry
lines and therefore protects the DPs in the correlated regime, while the Fu--Kane parity analysis
assigns nontrivial $Z_2$ indices to multiple hybridization gaps over nearby chemical potentials.
Nevertheless, a quantitative description of the correlation-driven renormalization and the detailed
evolution among the weak and strong TKI regimes and the DKSM as the chemical potential is tuned remain important directions for future study.
}
Equally crucial, experiments such as high-resolution ARPES, quantum oscillations, magnetotransport, and thermodynamic measurements could provide direct evidence, with expected signatures including nontrivial Berry phases, anomalous or topological Hall responses, and enhanced low-temperature specific heat, thereby establishing the phenomenology of topological heavy fermions.

\section*{Discussion}
Although the KFBs are anticipated to host strong correlation physics, observing such phenomena in most kagome systems has been challenging. This difficulty arises from the presence of many other dispersive conduction bands and the positioning of the FBs far away from $E_{\rm F}$. Interestingly, recent studies on Ni$_3$In, a $d$-electron kagome system characterized by a FB precisely located at $E_{\rm F}$, have demonstrated heavy fermion-like behavior despite the absence of local moments from $f$ electrons~\cite{Ye2024}. In this system, both essential components for the manifestation of heavy fermion behavior—localized moments and conduction electrons—are derived entirely from band features. The hopping-frustrated KFB acts analogously to the localized moments observed in conventional heavy fermion systems, providing a distinct mechanism for the emergence of heavy fermion-like phenomena and non-Fermi liquid behavior. \red{In contrast, YCG combines $f$-electron Kondo physics with kagome flat bands, thereby realizing topological heavy fermions rather than band-derived heavy fermion-like states.}

YCG is a unique system distinguished by the coexistence of two distinct FBs right near $E_{\rm F}$: (i) the KRSs and (ii) the KFB induced by hopping frustration (Figs.\,\ref{fig:kfb-krs}a-b). 
\ykcc{
The coexistence of the KRSs and the KFB not only enriches the correlated phase diagram but also yields filling-tunable weak and strong TKI regimes together with symmetry-protected Dirac crossings, thereby establishing YCG as a platform for topological heavy fermions.
}
Notably, the Cr  $ d_{z^2}$ orbital, which forms the KFB, exhibits hybridization with the Yb 4$f$ states opening the possibility of a unique interplay between the two different correlated FBs. When the KFB can be precisely located at $E_{\rm F}$, strong correlation physics can be induced as observed in Ni$_{3}$In system. As a result, the temperature scale of the system is suppressed and strange metallic behavior can occur. On the other hand, an increase in the density of states at $E_{\rm F}$ may enhance the Kondo hybridization and temperature scales. Therefore, it would be interesting to investigate experimentally how the two different energy scales arising from these two distinct FBs manifest. For instance, one can tune the KFB position by introducing strain~\cite{Kim2024} or doping to investigate the interplay between two distinct FBs near $E_{\rm F}$, 
\ykcc{thereby not only exploring the mutual influence of geometrically frustration-induced KFBs and
$f$-orbital KRSs but also providing experimental access to Dirac--Kondo crossings and to distinct weak and strong TKI regimes across nearby fillings, together with characteristic signatures of topological heavy fermions.}

\section*{Methods}

\section*{Crystal Growth}

Single crystals of YCG and LCG were synthesized using tin as a flux in sealed quartz tube under partial argon atmosphere. Yb (grain, 99.9\%), Lu (grain, 99.9\%), Cr (powder, 99.995\%), Ge (lump, 99.999\%), and Sn (shot, 99.995\%) were placed in an alumina crucible with an atomic ratio of Yb(or Lu):Cr:Ge:Sn = 1:3:6:20. The crucibles were sealed in quartz tubes under 5 Torr of Argon (99.9999\%) after purging 3--5 times. The ampules were heated to 1150 $^\circ$C and held for 24 hours, then cooled to 850 $^\circ$C at a rate of 250 $^\circ$C/hour. Further cooling to 600 $^\circ$C was performed at a rate of 1$^\circ$C/hour, at which point the Sn flux was removed by centrifugation. \red{Single crystal X-ray diffraction analysis of YCG single crystals confirmed full stoichiometry with absence of site disorder across all crystallographic sites.}

\section*{ARPES Measurements}

Angle-resolved photoemission spectroscopy (ARPES) measurements are conducted at beamline 21-ID-1 at the National Synchrotron Light Source II (NSLS-II) of Brookhaven National Laboratory (BNL) using a Scienta DA30 analyzer. Single-crystalline YCG samples are cleaved in an ultra-high vacuum with pressure below 10$^{-11}$ torr. $k_{x}$-$k_{y}$ constant-energy maps are measured at a temperature of 18~K using linearly horizontally (LH) polarized 93~eV light. To obtain a $k_z$-dependent constant-energy map, LH-polarized beams with photon energies ranging from 63 to 135~eV are used. All ARPES map data are acquired using the deflection mode of the Scienta DA30 analyzer. Temperature-dependent ARPES measurements are performed with an LH-polarized 97~eV beam, covering temperatures from 18~K to 220~K. 

\red{Single-crystalline LuCr$_6$Ge$_6$ samples are cleaved in an ultra-high vacuum with pressure below 10$^{-11}$ torr. $k_{x}$-$k_{y}$ constant-energy maps (See Supplementary Note 2.1) are measured at a temperature of 18~K using linearly horizontally (LH) polarized 200~eV light.}

\section*{Resonant photoemission spectroscopy measurements}
Resonant photoemission spectroscopy measurements are conducted at the 4A1-$\mu$ ARPES beamline of the Pohang Accelerator Laboratory (PAL) using LH-polarized light with photon energies ranging from 180~eV to 200~eV at a temperature of 80~K. Single-crystalline YCG samples are cleaved in an ultra-high vacuum with a pressure below 10$^{-10}$ torr.

\section*{DFT calculations}
We performed DFT calculations as implemented in the Vienna Ab initio Simulation Package (\textsc{VASP})~\cite{PhysRevB.54.11169} with pseudopotentials generated by the projector-augmented wave (PAW) method~\cite{PhysRevB.59.1758}. The generalized gradient approximation (GGA) of Perdew-Burke-Ernzerhof (PBE)~\cite{PhysRevLett.77.3865} was employed for the exchange-correlation energy functional. The kinetic energy cutoff for the plane-wave basis was set to 395~eV. \red{In the case of DFT+$U$ calculations, we employed the rotationally invariant formalism\cite{PhysRevB.57.1505}}. We used atomic parameters obtained from \red{single crystal X-ray analysis on our single crystalline samples.} Self-consistent field calculations were performed with a $\Gamma$-centered sampling of $k$-points on a $13 \times 13 \times 7$ uniform grid, using the tetrahedron method with Blöchl corrections~\cite{PhysRevB.49.16223}. The total energy was successfully converged below a threshold of $10^{-7}$~eV. To compute the irreducible representations of electronic states, we utilized IRVSP~\cite{GAO2021107760}. The VASPKIT~\cite{VASPKIT} code was used to plot the charge density of the wave functions. All band structure calculations included spin-orbit coupling.

\section*{DFT+DMFT Calculations}
We employed a fully charge self-consistent DFT+DMFT method implemented in DFT+Embedded DMFT (eDMFT) functional code~\cite{Haule2010} to describe the $f-c$ hybridization properly. DFT calculations were performed using the WIEN2k code~\cite{Blaha2020} and the DMFT loop treats the correlation effect of the Yb 4$f$ orbitals \red{and Cr 3$d$ orbitals}. The number of plane wave as set to $RK_{max}=9$ and a 2000 $K$ points set was used. The hybridization energy window from -10 to 10 eV was chosen with respect to the Fermi level. \red{We used $U = 7$ eV and $J_H = 0.7$ eV for the Yb 4$f$ orbitals, and $U = 5$ eV and $J_H = 0.7$ eV for the Cr 3$d$ orbitals.}. The local basis \red{of Yb atom} for the DMFT part is projected to $f_{5/2}$ and $f_{7/2}$ states so that the crystal field splittings are only considered on the level of lattice (DFT part) but are not normalized by the impurity solver. The continuous-time quantum Monte Carlo (CTQMC) solver was adopted to solve the impurity problem~\cite{Haule2007}. The nominal double-counting method was used, where the nominal occupancy of the Yb atom was set to 13\red{ and that of the Cr atoms was set to 3}.

\section*{Transport Measurements}
The AC resistance was measured using a Lakeshore Cryotronics 370 AC resistance bridge with a standard four-probe configuration, while temperature control was achieved using a Teslatron PT (Oxford Instruments). DC heat capacity measurements were performed using the relaxation method with a Physical Property Measurement System (PPMS, Quantum Design) and an SR850 Lock-In Amplifier (Stanford Research Instruments). 

\section*{Data availability}
The ARPES raw spectra, processed datasets, and calculation files generated in this study have been deposited in a public Zenodo repository and are available under DOI: 10.5281/zenodo.18712987. Additional data supporting this work that are not included in the public archive are available from the corresponding author upon request.

\section*{Acknowledgments}
This research was supported by Basic Science Research Program through the National Research Foundation of Korea (NRF) funded by the Ministry of Education (No. RS-2019-NR040081, RS-2023-00220471). 
Work at beamline 21-ID-1 at NSLS II was supported by the U.S. Department of Energy (DOE) Office of Science User Facility operated by Brookhaven National Laboratory under Contract No. DE-SC0012704. 
Y.K. acknowledges support from the National Research Foundation (NRF) of Korea, funded by the Ministry of Science and ICT (MSIT) (RS-2024-00456996 and No. RS-2019-NR040081), and the computational resources provided by the Korea Institute of Science and Technology Information (Grant No. KSC-2023-CRE-0382). 
B.G.J is supported by a grant from Kyung Hee University in
2023 (KHU-20233241), the Center for Advanced Computation (CAC) at Korea Institute for Advanced Study (KIAS), the National Supercomputing Center with supercomputing resources including technical support (KSC-2024-CRE-0131), and Center for Integrated Nanotechnologies, a DOE BES user facility, in partnership with the LANL Institutional Computing Program for computational resources.
Work at Yale supported by the Air Force Office of Scientific Research (AFOSR) under Grant No. FA9550-25-1-0026 and a BNL-Yale partner user agreement.
J.K. and C.K. are supported by the Global Research Development Center (GRDC) Cooperative Hub Program through the National Research Foundation of Korea (NRF) funded by the Ministry of Science and ICT (MSIT) (Grant No. RS-2023-00258359) and the NRF grant funded by the Korean government (MSIT) (Grant No. NRF-2022R1A3B1077234).
This research was supported by Sookmyung Women's University Research Grant No. 1-2403.2026.
This work was supported by the National Research Foundation of Korea (NRF) grant
funded by the Korea government (MSIT) (Grant No. RS-2025-00513951).
This research was supported by Basic Science Research Program through the
National Research Foundation of Korea(NRF) funded by the Ministry of Education(RS-2023-00248099). 
\red{This work was also supported by the Institute of Applied Physics, Seoul National University.}

\section*{Author Contributions}
H.L., C.L., B.G.J., Youngkuk K., B.S. and T.P. conceived the project;
H.L. and T.L. synthesized and characterized the materials with support from T.P.;
C.L. and Youngkuk K. conducted first-principles calculations and theoretically analyzed electronic structures;
J.K., S.L., and B.S. conducted ARPES measurements with support from A.R., A.K.K., E.V., C.K., C.H.A., and F.J.W.;
H.N. and B.S. analyzed resonant x-ray photoemission spectroscopy results;
H.N., J.K., B.S. conducted resonant x-ray photoemission spectroscopy with support from B.-G.P.;
B.S. analyzed ARPES results with support from J.K., Younsik K., and J.S.O.;
B.G.J. conducted dynamic mean-field theory calculations and theoretically analyzed electronic structures;
H.L., C.L., B.G.J., Youngkuk K., B.S. and T.P. wrote the paper with contributions from other authors;
All authors participated in the discussions and commented on the manuscript.

\section*{Competing Interests}
There is no competing interest.

\section*{References}
\bibliographystyle{naturemag}
\bibliography{refs}


\newpage

\clearpage

\clearpage

\begin{figure}
    \centering
    \includegraphics[width=0.9\textwidth]{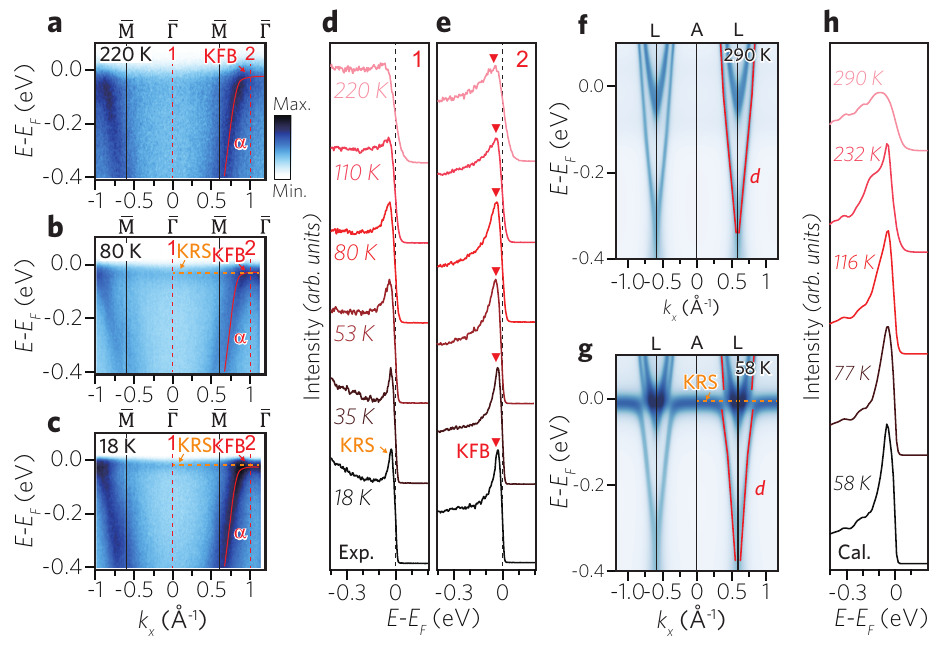}
    \caption{
    {\bf Temperature-dependent evolution of flat electronic structures.}
    (a-c) ARPES-measured E-$k$ cuts along the $\overline{\rm M}$-$\overline{\Gamma}$-$\overline{\rm M}$ high-symmetry line at temperatures of \red{220~K, 80~K, and 18~K}, respectively.
    (d,e) Temperature-dependent evolution of EDCs at $\overline{\Gamma}$ and $k_x$~=~1.0~$\AA^{-1}$.
    (f,g)  DFT+DMFT calculated \red{spectral functions} along the ${\rm L}$-${\rm A}$-${\rm L}$ high-symmetry line at temperatures of 58~K and 290~K, respectively. $d$ indicates Cr kagome bands.
    (h) Temperature-dependent evolution of the \red{spectral weight}, calculated using DFT+DMFT and integrated in momentum space with a Fermi-Dirac distribution.
    }
    \label{fig3}
\end{figure}

\clearpage

\begin{figure*}[t]
  \centering
  \includegraphics[width=\textwidth]{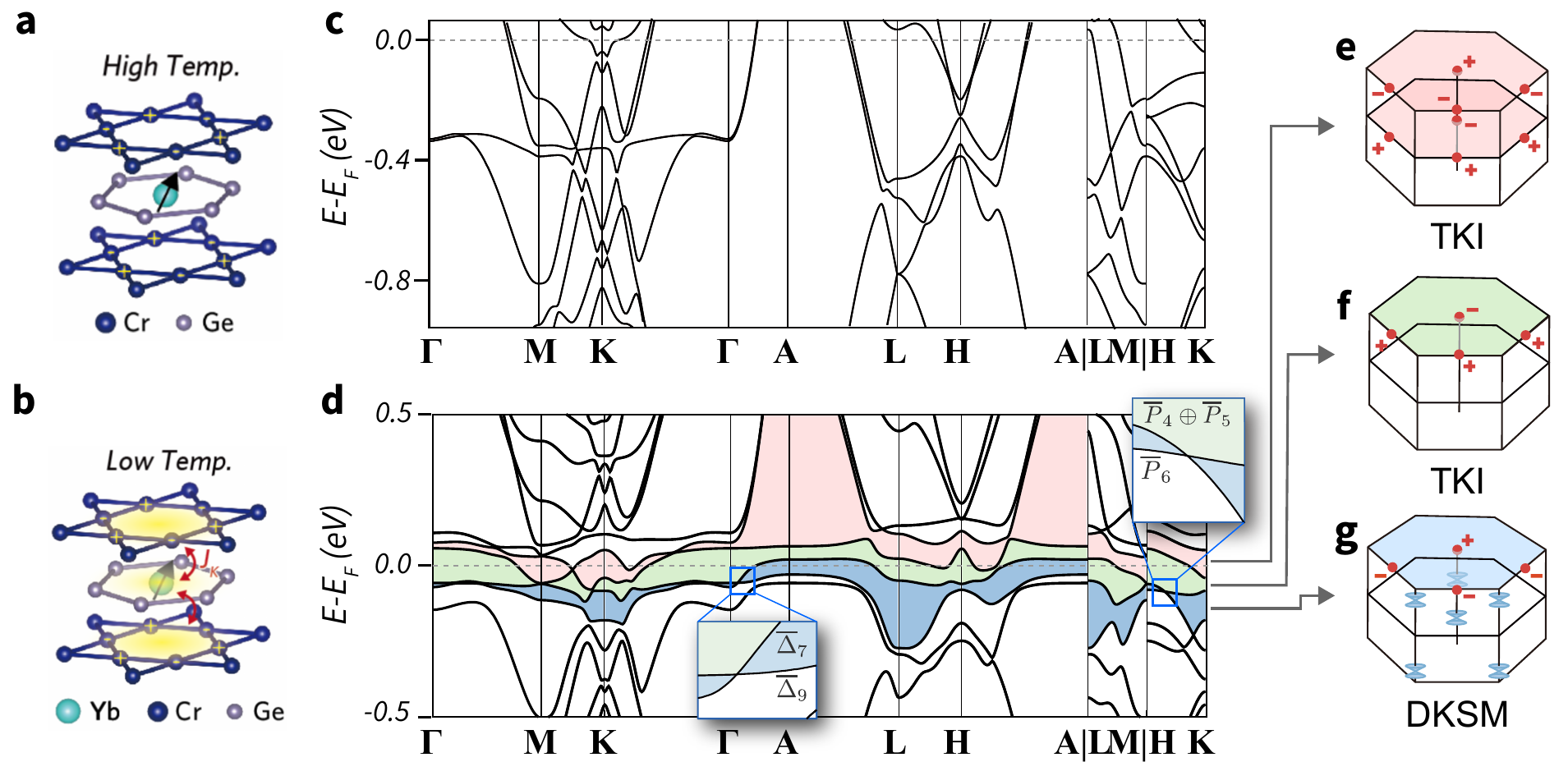}
  \caption{
(a,b) Schematics of the role of Yb in YCG: 
(a) high-temperature state with Cr (dark blue) and Ge (light purple) kagome layers and Yb (cyan); 
(b) Low-temperature state where yellow regions denote Dirac–Kondo fermions, localized in-plane but dispersive out-of-plane; red arrows indicate the Kondo coupling $J_K$.
\ykcc{
(c) DFT band structure of the high-temperature phase along the high-symmetry lines shown in Fig.\,2b.  The dashed line denotes the Fermi level $E_F$.
(d) DFT band structure of the low-temperature YCG phase computed with $U=2.6\,\mathrm{eV}$ for Cr $3d$ and $U=0\,\mathrm{eV}$ for Yb $4f$,  with the $U$ values and $E_F$ calibrated to reproduce the low-energy dispersion obtained from the preceding DFT+DMFT calculations, consistent with ARPES.  Compared with (c), hybridization opens multiple narrow gaps near $E_F$. 
Colored shading highlights three representative hybridization-gap regions, and arrows indicate their correspondence to the parity-based topological classification in (e--g).  Insets (blue boxes) enlarge the representative Dirac crossings. (e--g) Parity eigenvalues $(\pm)$ at the time-reversal-invariant momenta for the three colored gaps in (d), used to determine the Fu--Kane $Z_2$ indices. 
(e) weak topological Kondo insulator (TKI) with $(\nu_0;\nu_1\nu_2\nu_3)=(0;001)$, implying $\nu_{2\mathrm{D}}(k_z=0)=\nu_{2\mathrm{D}}(k_z=\pi)=1$. 
(f) strong TKI with $\nu_{2\mathrm{D}}(k_z=\pi)=1$ and $\nu_{2\mathrm{D}}(k_z=0)=0$. 
(g) Dirac--Kondo semimetal (DKSM), where the $k_z=\pi$ plane hosts a nontrivial $Z_2$ invariant while symmetry-protected DPs persist along the $\Gamma$--$A$ and $K$--$H$ lines, which prevents the opening of a full bulk gap. The blue Dirac cones indicate the locations of the DPs in the BZ.
}
}
\label{fig:kfb-krs}
\end{figure*}

\end{document}